\documentclass[aps,pra,letterpaper,twocolumn,superscriptaddress,floatfix]{revtex4-1}
\usepackage{amsmath,amssymb} 
\usepackage[utf8]{inputenc} 

\usepackage{graphicx} 
\usepackage{hyperref}
\hypersetup{colorlinks=true,citecolor={black},linkcolor={blue},urlcolor={blue}} 
\usepackage{xcolor}
\usepackage{bm}

\graphicspath{{.}{./fig/}}

\newcommand{\field}[1]{\mathbb{#1}} % requires amsfonts

\newcommand{\rme}{\mathrm{e}}
\newcommand{\rmi}{\mathrm{i}}

\renewcommand{\r}{\mathbf{r}}
\newcommand{\R}{\mathrm{R}}

\newcommand{\G}{\mathcal{G}}

\newcommand{\LTadd}{Institute for Theoretical Physics, University of Innsbruck, A--6020 Innsbruck, Austria}

\begin{document}

\title{Long Josephson junctions with exciton-polariton condensates}

\begin{abstract}
We demonstrate the possibility to build stable Josephson $\pi$-junction stripes with exciton-polariton condensates. 
The stability of the $\pi$-junction between arbitrary long polariton stripes is achieved at low pumping by balancing the snaking instability with counter-propagating flows towards the junction. Not dissimilar from a dark soliton, the instability becomes relevant at high pumping leading to formation of vortex dipoles. 
The resulting structures can be stabilised to produce static lattices of Josephson vortices in straight and ring geometries.
Our results build towards realization of quantum technological applications based on the Josephson effect at room temperature.
\end{abstract}

\author{A. Mu\~{n}oz Mateo}
%\email{A.M.Mateo@massey.ac.nz}
\affiliation{
	Departamento de F\'{i}sica, Facultad de Ciencias, Universidad de La 
	Laguna, E–-38200 La Laguna, Tenerife, Spain
}
\author{Y. G. Rubo}
%\email{ygr@ier.unam.mx}
\affiliation{
	Instituto de Energías Renovables, Universidad Nacional Autónoma de México, Temixco, Morelos, 62580, Mexico
}
\author{L. A. Toikka}
%\email{lauri.toikka@gmail.com}
\affiliation{\LTadd}

\date{\today}
\maketitle

\emph{Introduction.---}Observation~\cite{Kasprzak06,Balili07,Lai07} of Bose-Einstein condensates (BECs) of exciton-polaritons (polaritons) in semiconductor microcavities~\cite{Kavokin17lib}, including formation of macroscopic quantum coherence at room temperature~\cite{Baumberg08,Lerario17,Kang19}, offer a new platform for realizing quantum technological applications. The driven-dissipative polariton superfluids, however, are essentially different from their equilibrium counterparts such as liquid helium or ultra-cold atoms. 
Apart from the composite character of polaritons as mixed states of excitons and photons, and their finite lifetime in microcavity, a crucial difference is in the presence of condensate currents even in steady states. 
These currents usually originate from non-uniform pumping that is necessary to excite the polariton BEC, but they can appear spontaneously as well~\cite{Nalitov17}. When the pump-dissipation balance is nonlocal, the flows of particles appear from regions of positive to negative balance~\cite{Keeling2008}. The polariton supercurrents can lead, for instance, to the generation of quantized vortices (loop currents) due to inhomogeneities of the samples~\cite{Lagoudakis2008}. 

The steady-state polariton flows play a fundamental role in the distinct dynamics of topological defects in polariton condensates. The steady-state currents depend sensitively on the particular geometry of the polariton pumping scheme, which in turn can lock the resulting stationary configuration of vortex lattices~\cite{Tosi2012}. Even low-intensity flows are able to drag a single vortex out of the system~\cite{Keeling2008,Sun19}, and more complicated arrangements have to be designed to achieve the dynamical stability of these defects~\cite{Ma2018}.  The design of excitation scheme is also of key importance in recently proposed polariton simulators and graphs~\cite{Berloff17,Lagoudakis17}. Underpinning the locking of relative phases, the basic element of polariton simulators and networks is the Josephson junction of two condensates~\cite{Wouters08,Eastham08}. The Josephson junctions are frequently formed with a phase difference of $\pi$ between the condensates~\cite{Read10,Manni11,Ohadi16}. In this case, the order parameter has a node in between the small-sized condensates, while the dark soliton, or 1D curve of zero density, should be imprinted for large-sized condensates forming a long Josephson junction (LJJ). 
LJJs are significant in their ability to host Josephson vortices~\cite{Kaurov05} that are strictly localised within the junction. The long lifetime and form-stability of dark solitons in equilibrium BECs can be exploited for qubit operations by using the soliton profile as a non-harmonic potential for a two-level system~\cite{Shaukat17}. 
LJJs in polariton BECs offer exciting prospects for realizing the ac Josephson effect~\cite{Lagoudakis10} 
% and highly sensitive quantum magnetic field detectors (SQUIDs)
% YR: I commented this because polaritons, in the first approximation, do not have charge, so that the magnetic field detection is questionable.  
at room temperature without the need for cryogenic refrigeration.

In spite of the vast technological importance, the emergence and stability of LJJs with polariton condensates is not comprehensively understood. In an incoherently pumped homogeneous polariton BEC, a stationary dark soliton (LJJ) is always unstable to any small perturbation, which leads to acceleration and blending of the dark soliton with the background~\cite{Smirnov14}. 
In a one-dimensional (1D) waveguide geometry,  experimental realization of dark solitons in polariton fluids has been achieved recently~\cite{Walker2017}. 
More recently, Josephson vortices have been experimentally observed to emerge from a phase twist imposed at the boundary of a polariton condensate~\cite{Caputo2019}. An imprinted $\pi$ phase difference between resonant lasers beams situated at two spatially separated spots was observed to produce at low pumping power a robust LJJ in the phase of a two-dimensional (2D) polariton cloud. 
Further increase of the non-resonant-pump power (and, equivalently, of the polariton density) was observed to lead to the decay of the domain wall into stable vortex dipoles, in a similar way as Josephson vortices (or fluxons) emerge in LJJs between superconductors in the presence of external magnetic fields~\cite{Tinkham2004}. 

In this Letter, we demonstrate how it is possible to form arbitrarily long stable Josephson junctions (dark solitons) and the vortex chains on demand, by exploiting the out-of-equilibrium nature of polariton condensates and the steady-state currents. 
We show that, contrary to their equilibrium counterparts, and contrary to the homogeneous polariton condensate, there is a parameter regime where the dark soliton between two stripe condensates is stable. The snaking instability is suppressed by the counter-propagating flows towards the soliton line. The vortex instability remains important nonetheless, and it can be initiated at certain range of parameters and excitation conditions, leading to formation of stable 1D lattices of vortex-antivortex dipoles.

\begin{figure}[tb]
	\centering
	\includegraphics[width=\linewidth]{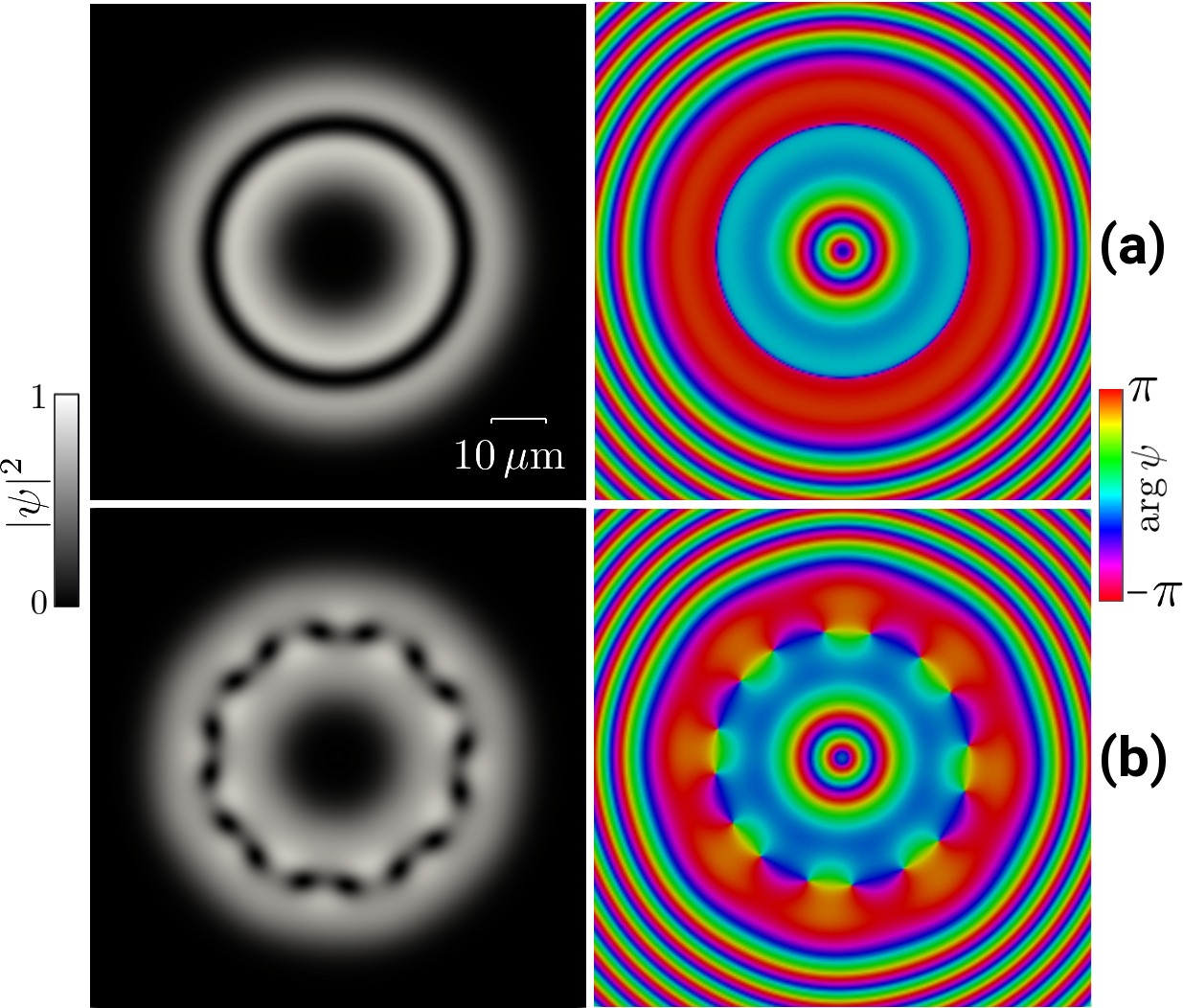}
	\caption{
	(a) Ring-shaped long Josephson junction with imprinted stable dark soliton, and
	(b) formation of stable vortex-dipole train. 
	Both configurations are generated by two concentric Gaussian pumps of radial width 
	$\sigma=7.7\,\mu$m, separated by a radial distance $d=20\,\mu$m, and situated equidistantly from a middle circle of radius $21.5\,\mu$m. 
	The pumping intensities are (a) ${\hbar}P_0=58\,\mathrm{meV}/\mu\mathrm{m}^2$ and (b) ${\hbar}P_0=59\,\mathrm{meV}/\mu\mathrm{m}^2$. 
	The lifetime of states in (a) and (b) exceeds $1~\mathrm{ns}$, and the vortex chain in (b) is stable with the vortices locked in place.}
	\label{fig:ring}
\end{figure}

\emph{Model.---}To incorporate the dynamical balance of pumping and loss, we use a coupled mean-field description for the macroscopic wavefunction of the BEC $\psi(\r,t)$ and density of the reservoir polaritons $n_\R(\r,t)$,
\begin{subequations}
	\label{eqn:GPE-sys}
	\begin{align}
	\label{eqn:GPE-sys-1}
	\rmi \hbar \frac{\partial \psi}{\partial t} &= \left \lbrace - \frac{\hbar^2}{2m}\nabla^2  + g\left| \psi \right|^2 + \left(g_\R + \rmi \frac{\hbar r}{2} \right) n_\R-\rmi \frac{\hbar \gamma}{2}  \right \rbrace \psi,\\
	\label{eqn:GPE-sys-2}
	\frac{\partial n_\R}{\partial t} &= P - \left(\gamma_\R + r \left| \psi \right|^2  \right) n_\R,
	%- D \nabla^2 \right) n_\R,
	\end{align}
\end{subequations}
where the rate of amplification $r n_\R$ of the condensate from reservoir-induced stimulated scattering is linear in the reservoir density. Here $m$ is the effective mass of polaritons, $g > 0$ represents polariton-polariton interactions, $g_\R$ represents interactions between the reservoir polaritons and condensate polaritons, $\gamma$ defines the loss rate of polaritons from microcavity, $P$ is the pumping rate of polaritons into the reservoir, and $\gamma_\R$ is the loss rate of reservoir polaritons.

We consider condensate excitation in ring geometry, which attracts much interest recently, and 
assume that the pumping consists of two static concentric rings with a Gaussian profile, a spacing of $d$ and width $\sigma$: 
$P(r)= P_0[\G_-(r) + \G_+(r)]$, where $\G_\pm(r) = \exp\left\{-\frac{(r\pm d/2)^2}{2\sigma^2}\right\}$.
This excitation produces two polariton condensates with the azimuthal Josephson junction. In conservative systems \cite{Tinkham2004}, this setting supports Josephson-vortex solutions that produce localized loop currents centred at the junction. While the same profiles can exist in the polariton BEC as well, we focus here on the symmetry conserving azimuthally uniform condensates with an imprinted long dark soliton between them (Figs.\ \ref{fig:ring}(a,b)). 

To get more insight into the stability of the long dark solitons, we consider the limit of large radii, which is equivalent to the case of two parallel pumping stripes in the $y$-direction: $P(x)=P_0[\G_-(x) + \G_+(x)]$.
% with constants $P_\pm$ describing possible asymmetry of the stripes. YR: We probably will not consider asymmetry below.
Under such pumping, Eq.~\eqref{eqn:GPE-sys} can be written as Josephson-type equations for the two parts of the system. Specifically, coordinate-separable solutions for Eq.~\eqref{eqn:GPE-sys} can be approximated by $\psi(x,y; t)  = \phi_+ (y,t)\, {\G}_+^{\frac{\alpha - \rmi \beta}{2}}(x) + \phi_-(y,t) \, {\G}_-^{\frac{\alpha - \rmi \beta}{2}}(x)$ and $n_\R(x,y;t) = n_+(y,t)\, \G_+(x) + n_-(y,t)\, \G_-(x)$, where we model the condensate width and flow towards $x = 0$ through the real constants $\alpha$ and $\beta$. Substitution to system~\eqref{eqn:GPE-sys} gives
\begin{subequations}
	\label{eqn:TwoMode-Josephson-simpl}
	\begin{align}
	\label{eqn:GPE-TwoMode-Josephson-simpl}
	\rmi \hbar  \frac{\partial \phi_\pm}{\partial t}   
	&= \left(\Gamma -\frac{\hbar^2 \partial_y^2}{2m}+\tilde g\left| \phi_\pm \right|^2 + \Gamma_\R\,n_\pm \right)\phi_\pm - J  \phi_\mp  , \\
	\label{eqn:Res-TwoMode-Josephson-simpl}
	\frac{\partial n_\pm}{\partial t} &=  P_0 - \left( \gamma_\R + \tilde r  \left| \phi_\pm \right|^2 \right) n_\pm ,
	\end{align}
\end{subequations}
where the coefficients $ \Gamma,\, \Gamma_\R,\, J\, \in \field{C}$ and $\tilde g,\, \tilde r, \, \in \field{R}$ are explicitly given in the Supplementary Material. We note that the Josephson coupling is, in general, complex, where a positive imaginary part of $J$ stems from the overlap of the ring wave functions. Distinct from the Josephson coupling characterised by the real part of $J$, it describes the dissipative coupling of the two rings. Presence of dissipative coupling favors the formation of $\pi$ phase difference between the rings at low pumping, thus imprinting the dark soliton between them \cite{Aleiner2012}, see Fig.\ \ref{fig:ring}(a).

\emph{Long Josephson junctions.---}The resulting system \eqref{eqn:TwoMode-Josephson-simpl} describes a long Josephson junction along the $y$-direction characterised by the complex coupling $J$. We begin by demonstrating that the vortices in this system emerge from dynamical instabilities at the nodes of transverse standing waves excited on the low-density lines that configure the junction. 
To show this we focus on the symmetric $\phi_+=\phi_-$ and anti-symmetric $\phi_+= -\phi_-$ solutions to Eq.~\eqref{eqn:TwoMode-Josephson-simpl} without transverse variation, that is $\phi_+(y,t)=\sqrt{n_0} \exp(-\rmi \mu t)$ and $n_\pm=n_{R0}$. 
We now consider the corresponding generalized Bogoliubov equations \cite{Wouters2007} 
for the linear excitations $\mathbf{u}_\pm(y)=[u_\pm(y), v_\pm(y),w_\pm(y)]$ of these states, and write $\phi_\pm(y; t) \rightarrow   \{ \sqrt{n_0}   + u_\pm(y)  \exp(-\rmi \omega t) + v_\pm^*(y)  \exp(\rmi \omega^* t)  \}  \rme^{-\rmi \mu  t}$ and $n_\pm(y,t)\rightarrow \{n_{R0} + w_\pm(y) \exp(-\rmi \omega t) + w_\pm^*(y)   \exp(\rmi \omega^* t)\} $. The excitation spectrum can be readily solved by Fourier-basis expansion $\mathbf{u}_\pm(y)=\mathbf{u}_{k,\pm} \exp(\rmi k  y)$. The resulting dispersion presents four energy branches parametrised by the
Rabi frequency $\omega_J=2|\mathrm{Re}[J]|/\hbar$, which is induced by the coupling $J$. 
For the experimentally relevant regime $\gamma/\gamma_\R\ll 1$, the stability spectrum can be approximated by
 \begin{align}
  \omega=-\rmi\frac{\eta}{2}\pm\sqrt{\left(\frac{\hbar k^2}{2m} \pm \omega_J\right)\left(\frac{\hbar k^2}{2m}+
  	\frac{2gn}{\hbar}\pm \omega_J\right)-\frac{\eta^2}{4}},
 \end{align}
where $\eta=\gamma/(1+\gamma_\R n_\R/\gamma n)$, and the $\pm$ sign inside the square root accounts for the symmetric (--) and anti-symmetric (+) states. As can be seen, analogously to conservative systems, the contribution of $-\omega_J$ to the dispersion involves dynamical instabilities responsible for the production of vortices either for the anti-symmetric state when $\mathrm{Re}[J]>0$, or for the symmetric state when $\mathrm{Re}[J]<0$.

\begin{figure}[tb]
	\centering
	\includegraphics[width=\linewidth]{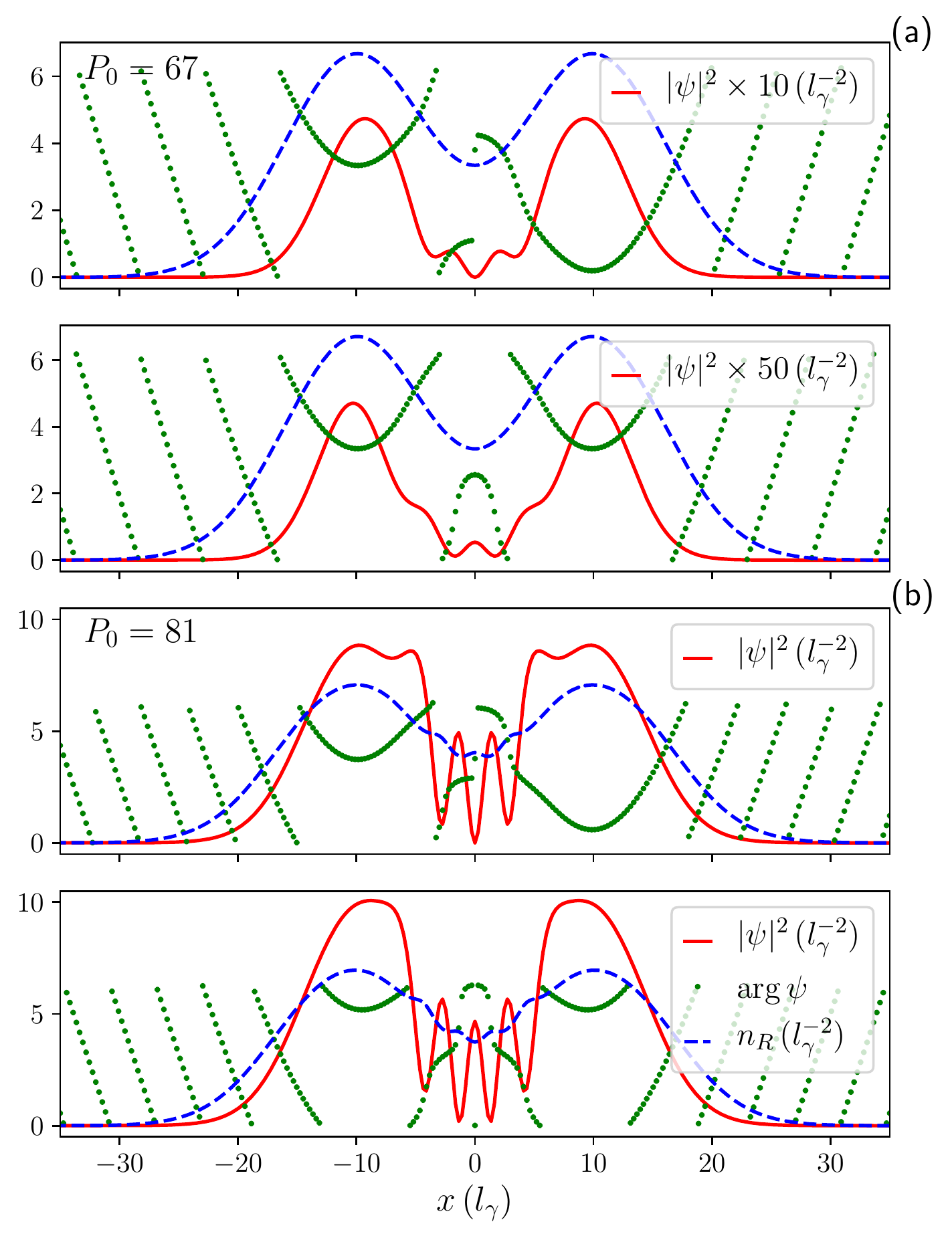}
	\caption{Anti-symmetric (top) and symmetric (bottom) states in a double Gaussian-stripe at low pumping $P_0=67$, panels (a), and intermediate pumping $P_0=81$, panels (b). In both cases, the anti-symmetric states are dynamically stable, whereas the symmetric states are unstable.}
	\label{fig:sol_lowP}
\end{figure}
\begin{figure}[tb]
	\centering
	\includegraphics[width=\linewidth]{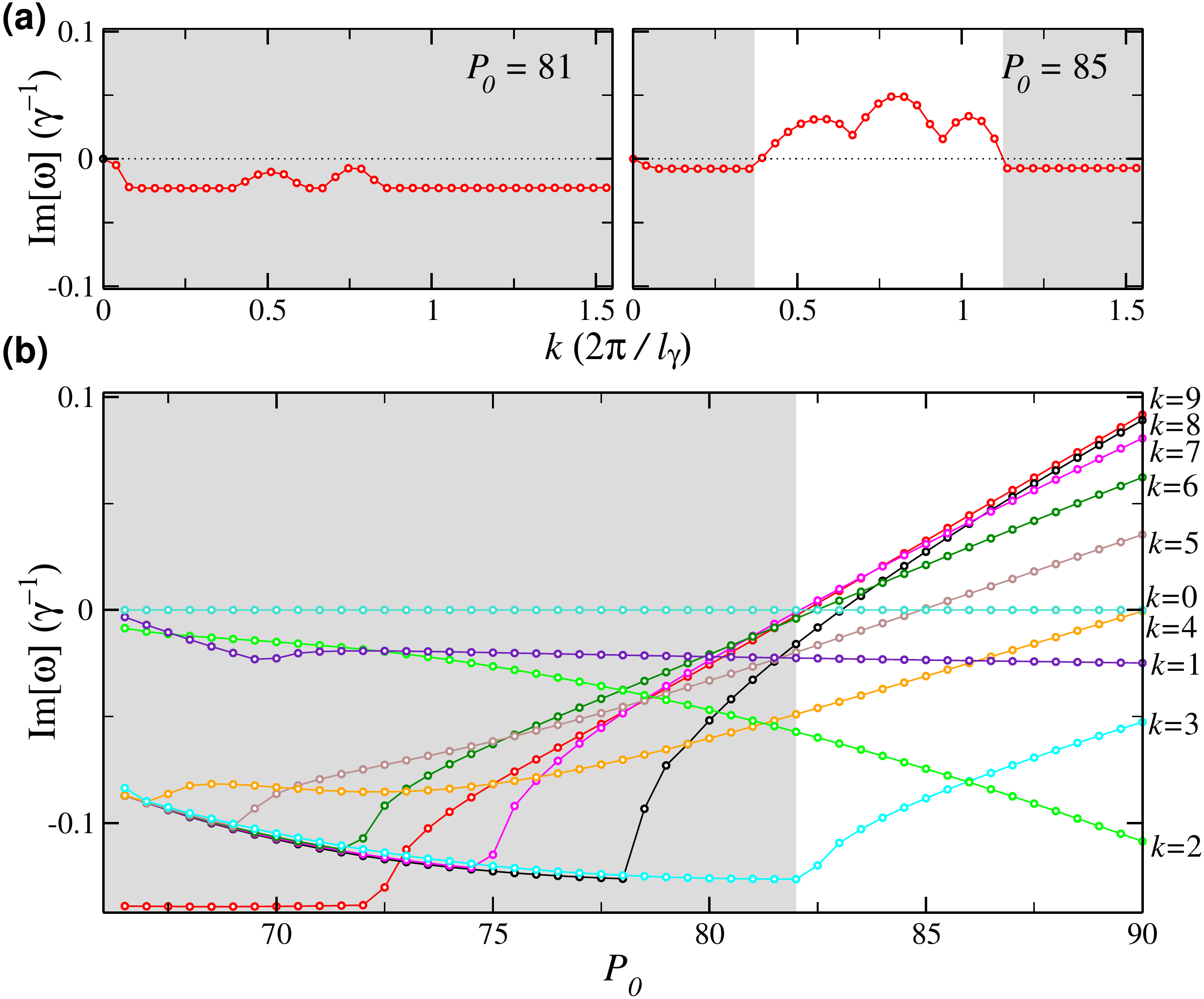}
	\caption{Imaginary part of the linear excitation frequencies around stationary anti-symmetric states in a double Gaussian-stripe pump, for fixed pumping intensities [panels (a)] and fixed transverse domain length $L_y=80 \,l_\gamma$ [panel (b)]. In the latter case, different curves correspond to different values of the transverse wavenumber $k$ (in units of $2\pi/L_y$), as labeled on the right margin. Left panel of (a) corresponds to Fig.~\ref{fig:sol_lowP}(b). The shaded regions indicate  dynamical stability of the anti-symmetric state.
	}
	\label{fig:bog_s6R10a}
\end{figure}

For comparison with experimental parameters, we set $\hbar\gamma = 1$ meV, as energy unit, and $\gamma^{-1}=0.66$ ps and $l_\gamma=\sqrt{\hbar/m\gamma}=0.77\,\mu$m as time and length scales. In these units, we use  $g = 4\pi\times0.0054$, $g_\R = 0.085$, $r = 0.17$, and $\hbar\gamma_\R = 10$ in all our numerical calculations. Implementing the double Gaussian-stripe pump we choose $d=20$ and $\sigma=6$, and consider a range of pumping intensities $P_0\in[65,90]$ above the threshold for the generation of the polariton condensate. For homogeneous pumping the threshold is $P_\mathrm{th}=\gamma\gamma_\R/r \approx 58.83$. Figure \ref{fig:sol_lowP} depicts the steady states at low $P_0=67$ [panels (a)] and intermediate $P_0=81$ [panels (b)] pumping intensities. The anti-symmetric states [top panels in (a) and (b)] present a nodal line at $x=0$ and will be referred to as dark solitons. Both the symmetric states (bottom panels) and generally the anti-symmetric states show off-center density dips associated with steep phase gradients. These shoulders are situated symmetrically around $x=0$, and will be referred to as gray solitons due to the non-zero background polariton current at their locations. As it can be seen, the gray solitons develop deeper density depletions for increasing pumping. For the considered ratio $2\sigma/d=0.6$, the analysis of linear excitations (Fig.~\ref{fig:bog_s6R10a}) shows that the anti-symmetric states are dynamically stable in the range $P_0\in[65,82]$, whereas the symmetric states are unstable. 
This happens in spite of the fact that, for given pumping, both configurations alternate in hosting the maximum number of particles. As was shown in Ref.~\cite{Ohadi16}, the most populated state (either symmetric or anti-symmetric) follows a periodic variation with the oscillation frequency $\mu$, hence with the pumping intensity $P_0$, which is well approximated by the sign of $\cos(\sqrt{2m\mu/\hbar}\,d-\pi/4)$,  positive when the symmetric state is the most populated and negative otherwise. In the shaded region of Fig.~\ref{fig:bog_s6R10a}, the most populated state can be the symmetric one, while only the anti-symmetric state is stable. 
%While unstable, the strength of the instability of the most highly populated symmetric state can be small.

Figure \ref{fig:bog_s6R10a} collects our results for the linear stability spectrum from the numerical solution of the Bogoliubov equations in a periodic domain of length $L_y$ along the transverse coordinate. The imaginary part of the frequencies of linear excitations around the stationary anti-symmetric states is plotted against the modulus of transverse wavenumber $k$, for two pumping cases [panels (a)], and against the intensity of pumping, for fixed domain length $L_y=80\,l_\gamma$ [panel (b)]. In the latter, the different curves correspond to the maximum imaginary frequency for given values of $k\in[1,9]\times2\pi/L_y$; higher wavenumbers do not make any difference regarding stability.
As can be seen, dynamically stable dark solitons can be found in 2D polariton fluids at low and intermediate pumping. 
The  distinctive feature of these solitons compared to their conservative counterparts is that here there is no limitation for the length $L_y$ of the transverse dimension. In equilibrium BECs, dark solitons are unstable against bending of the soliton stripes that leads to the appearance of vortices. This so-called snaking instability~\cite{PhysRevA.87.043601} is a long-wave-length phenomenon that can be only prevented if the transverse size of the system is less than a threshold length, typically a few healing lengths $\xi=\hbar/\sqrt{m g n}$, where $n$ is the condensate density and $g$ the interaction strength \cite{KuznetsovTuritsyn1988}. As a consequence, only relatively small 2D systems are capable of supporting stable solitons in equilibrium condensates. On the contrary, the stability of the 2D solitons presented here is not sensitive to the transverse length of the system, and an arbitrarily long Josephson $\pi$-junction can be stable. The mechanism of the stability is dynamical, and resides on the existence of incoming currents towards the soliton notch as we elaborate below.

\begin{figure}[tb]
	\centering
	\includegraphics[width=\linewidth]{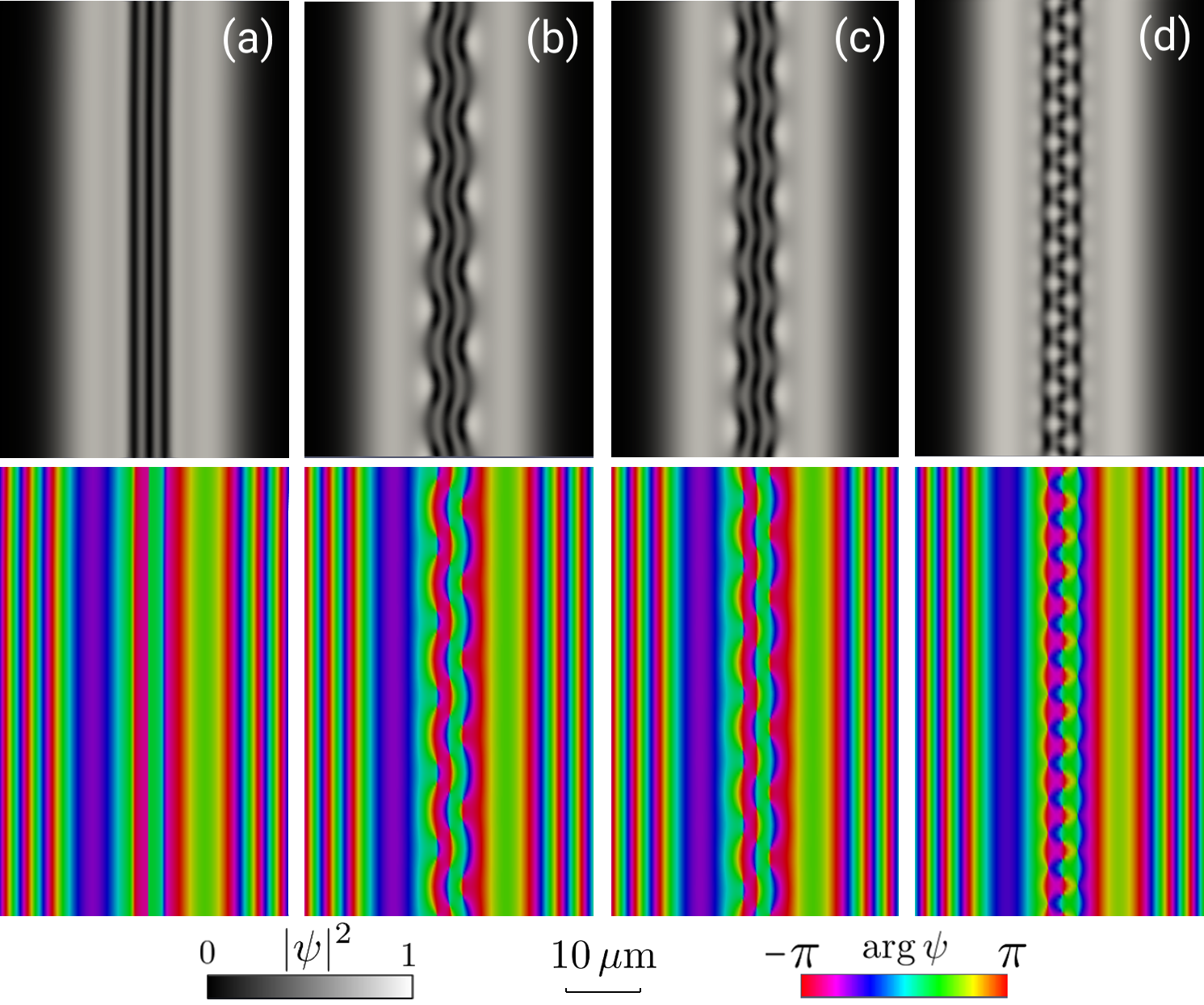}
	\caption{Vortex-dipole and solitonic lump configurations [(b)-(d)] obtained from the decay of dynamically unstable anti-symmetric stationary states (a).  Steady condensate densities (top panels) and phases (bottom panels) are shown. States (b) and (c) emerge from (a) at $P_0=85$, while state (d) is achieved at $P_0=90$. In all cases, the pump is a double-stripe Gaussian with $2\sigma/d=0.6$. 
	}
	\label{fig:p85_decay}
\end{figure}

The snaking instability can produce the decay of solitons in polariton fluids, and hence the appearance of vortices, if the pumping is high enough (non-shaded range in Fig. \ref{fig:bog_s6R10a}). However,  the character of the instability is different from that of equilibrium systems. As depicted in the top right panel of Fig. \ref{fig:bog_s6R10a}(a), for $P_0=85$, the unstable modes present both a maximum and a minimum wavenumber, whereas there is no such minimum in equilibrium BECs. The maximum unstable wavenumber $k_\mathrm{max}$ allows for preventing the snaking instability by choosing a short enough system with $L_y<k_\mathrm{max}^{-1}$. The minimum unstable wavenumber, on the other hand, constraints the type of bifurcated vortex states and so the minimum number of vortices emerging from the solitons. 

Our numerical results show that the unstable modes are exponentially localized along the axial $x$-direction, around the lowest density lines of the solitons, and excite standing waves along the transverse direction that depend on the system geometry. During the soliton decay, the vortices can be seen to emerge from the nodal points of the unstable standing waves producing $N_\mathrm{v}=2 k L_y$ vortices ($N_\mathrm{v}/2$ vortex dipoles). In fact, the emerging stable configurations comprise of vortex dipoles aligned along the original solitons. Figure \ref{fig:ring} shows a neat example of 8 vortex dipoles uniformly distributed on the initial position of a single ring dark soliton (without off-center solitons in this case for $2\sigma/d=0.77$). Analogously, Figure \ref{fig:p85_decay}(b) depicts the resulting configuration from the decay of a straight anti-symmetric state at $P_0=85$ (corresponding to the top right panel of Fig.~\ref{fig:bog_s6R10a}) after seeding a perturbation with $k=6\times 2\pi/L_y$. The vortex-dipole cores cannot be clearly discerned in this case, and the emerged low density waves are better described as solitonic lumps, or also as the Jones-Robert solitons \cite{Jones1982}. In both cases, either vortex dipoles or solitonic lumps, the inward currents towards the junction permit the static arrangement of these otherwise moving non-linear waves, which fly away from the junction in conservative BECs (see e.g. Ref.~\cite{Becker2013}). 

Multistability is another consequence of the combination of inward currents and multiple unstable channels. The appearance of unstable purely imaginary modes as a function of increased pumping (Fig.~\ref{fig:bog_s6R10a}) 
%, by crossing the horizontal axis (Im$[\omega]=0$) in Fig.~\ref{fig:bog_s6R10a} and getting a pure imaginary frequency since then, 
is associated with the bifurcation of a new family of stationary vortex states inheriting the nodal configuration of the unstable mode. Although concerning bifurcations the scenario is analogous to its conservative counterpart \cite{Chladni2014}, a crucial difference is the existence of multiple dynamically stable configurations between the new vortex states; a situation that manifests more clearly with strong pumping with more instability channels. For instance, Fig. \ref{fig:p85_decay}(c) shows another stable pattern of lumps from the decay of the anti-symmetric state at $P_0=85$, in this occasion as a result of feeding a perturbation with $k=7\times 2\pi/L_y$. At higher pumping, more complex configurations arise combining lumps and vortex dipoles, as can be seen in Fig. \ref{fig:p85_decay}(d) for $P_0=90$.

\emph{Conclusions.---}We have shown how arbitrarily long, dynamically stable Josephson junctions (dark solitons) can be generated on-demand in exciton-polariton semiconductor microcavities. Decay of the junction into a stable locked array of vortex dipoles can also take place at higher pumping intensity. Out of equilibrium, the dynamics and stability of the Josephson junction and the associated Josephson vortices are thus under complete control. A prominent outlook of our results concerns the dynamics of a linear LJJ array~\cite{Gil_Granados_2019}, and the implementation of dynamically stable weak links in ring geometry.

\emph{Acknowledgements.---}LT is supported by the Austrian Academy of Sciences (P7050-029-011). YR acknowledges support by CONACYT (Mexico) Grant No.\ 251808 and by PAPIIT-UNAM Grant No.\ IN106320.

\bibliographystyle{apsrev4-1}
\bibliography{polariton_soliton_vortex}

\end{document}